\documentclass[3p,sort&compress,twocolumn]{elsarticle}

\usepackage{amsmath}
\usepackage{lineno}
\usepackage{hyperref}
\usepackage{graphicx}
\usepackage{natbib}
\usepackage{subcaption}
\usepackage{siunitx}
\DeclareSIUnit\year{a}
\DeclareSIUnit\wavenumber{\per\cm}
\usepackage{placeins}
\usepackage[version=4]{mhchem}
\usepackage{booktabs}
\usepackage{multirow}
\modulolinenumbers[5]

\begin{document}

\begin{frontmatter}

\title{Highly efficient isotope separation and ion implantation of $^{163}$Ho for the ECHo project}

\author[kch-mainz,physik-mainz]{Tom Kieck\corref{correspondingauthor}}
\cortext[correspondingauthor]{Corresponding author} \ead{tomkieck@uni-mainz.de}

\author[kch-mainz]{Holger Dorrer}
\author[kch-mainz,gsi,him]{Christoph E. D\"{u}llmann}
\author[physik-mainz]{Vadim Gadelshin}
\author[kch-mainz]{Fabian Schneider}
\author[physik-mainz]{Klaus Wendt}

\address[kch-mainz]{Institute of Nuclear Chemistry, Johannes Gutenberg University, 55099 Mainz, Germany} 
\address[physik-mainz]{Institute of Physics, Johannes Gutenberg University, 55099 Mainz, Germany} 
\address[gsi]{GSI Helmholtzzentrum f\"{u}r Schwerionenforschung GmbH, 64291 Darmstadt, Germany} 
\address[him]{Helmholtz-Institut Mainz, 55099 Mainz, Germany}

\begin{abstract}
The effective electron neutrino mass measurement at the ECHo experiment requires high purity $^{163}$Ho, which is ion implanted into detector absorbers. To meet the project specifications in efficiency and purity, the entire process chain of ionization, isotope separation, and implantation of $^{163}$Ho was optimized. A new two-step resonant laser ionization scheme was established at the \SI{30}{\kilo\volt} magnetic mass separator RISIKO. This achieved ionization and separation efficiencies with an average of \SI[parse-numbers=false]{69(5)\textsubscript{stat}(4)\textsubscript{sys}}{\percent} using intra-cavity frequency doubled Ti:sapphire lasers. The implantation of a $^{166\textrm{m}}$Ho impurity is suppressed about five orders of magnitude by the mass separation. A dedicated implantation stage with focusing and scanning capability enhances the geometric implantation efficiency into the ECHo detectors to \SI{20(2)}{\percent}.
\end{abstract}


\end{frontmatter}


\section{Introduction}

 The ECHo \cite{Gastaldo2017} (Electron Capture in $^{163}$Ho) project is designated for measuring the electron neutrino mass by recording the decay spectrum following electron capture of $^{163}$Ho implanted in large arrays of Metallic Magnetic Calorimeters (MMCs) \cite{Gastaldo2013}. For this purpose the synthetic radioisotope $^{163}$Ho with a half-life of \SI{4570(50)}{\year} \cite{Baisden1983} is produced by intense neutron irradiation of an enriched $^{162}$Er target in the ILL high flux nuclear reactor. The Ho is then chemically separated from the target material and any other interfering elements \cite{Dorrer2018}. The obtained samples mainly consists of $^{163}$Ho, $^{165}$Ho (stable) and $^{166\textrm{m}}$Ho (radioactive). The latter is a long-lived nuclear isomer ($T_{1/2} = \SI{1132(4)}{\year}$ \cite{Nedjadi2012}). If present in the MMCs, the emitted radiation leads to undesired background in the spectrum. Therefore, it has to be separated before incorporation of the sample into the absorbers. The most convenient method is to mass-separate the ionized Ho sample.
 
\begin{figure*}[h]
	\centering
	\includegraphics[width=\textwidth]{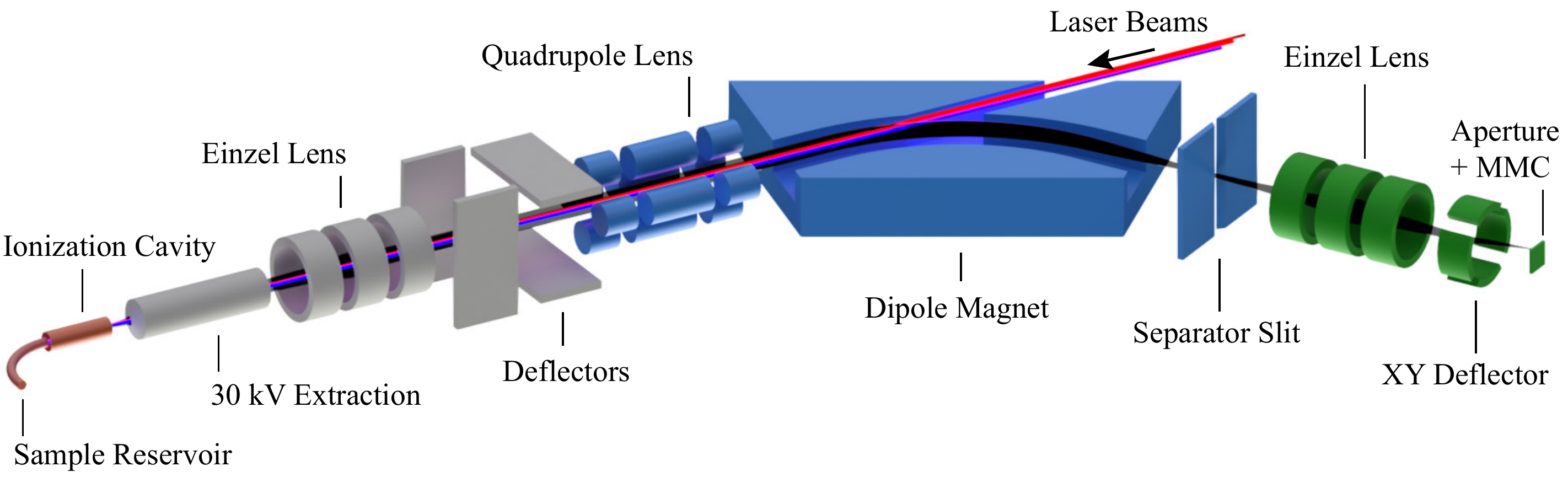}
	\caption{Overview of the RISIKO mass separator. The ion source development is described in Section \ref{sec:resonance}, the magnetic mass separation in Section \ref{sec:separation}, and the implantation into the metallic magnetic calorimeters (MMCs) in Section \ref{sec:implantation}.}
	\label{fig:risiko}
\end{figure*}

To obtain high efficiencies of the ionization and separation process, a resonance ionization laser ion source (RILIS) is used at the RISIKO isotope separator \cite{Zimmer1990} at Johannes Gutenberg University Mainz. It provides elemental selectivity in the formation of a high quality ion beam. A sketch of the major components is provided in Figure \ref{fig:risiko}. Similar combinations of a RILIS with a mass separator are commonly used worldwide at isotope separation on-line (ISOL) facilities for the efficient production and delivery of isotopically and isobarically pure ion beams of rare isotopes, which are typically produced by proton bombardment in a nuclear reaction target situated immediately at the ion source \cite{Letokhov1979,Koster2003,Fedosseev2017}.

For future phases of the ECHo experiment large amounts of $^{163}$Ho of more than \num{1e17} atoms enclosed in more than \num{e5} detectors are intended to deliver adequate statistics in the decay spectrum in a reasonable time frame \cite{Gastaldo2017}. Because of the limited availability of the primary source material and irradiation time at high flux nuclear reactors, an excellent efficiency of the ion source and implantation process as well as transmission through the separator is required. For this purpose the RILIS of the RISIKO mass separator is specifically optimized for Ho gas formation and ionization with minimum losses \cite{Kieck2018}.

In general, multi-step resonant laser ionization is a highly element-selective process leading to substantial reduction of unwanted elemental components in the ion beam. The ultimate limitation of selectivity is given by thermal ionization processes occurring on the hot surfaces \cite{Kirchner1990} that are necessary to produce and sustain single-atom vapor. Correspondingly this the influence of surface-ionization has to be characterized individually for each elemental composition in the ion source.

\section{Resonance ionization of Ho}
\label{sec:resonance}
\subsection{Sample material and atomization}
\label{sec:material}

All systematic measurements on stable $^{165}$Ho were performed using commercially available atomic absorption standard (AAS) solution (\SI{99.9}{\percent}) with a concentration of \SI{1}{\gram\per\liter}. The solution was diluted by a factor of ten with distilled water. Individual drops (\SI{3}{\micro\liter} volume, \SI{2}{\percent} volume accuracy) were dripped onto $\SI{3x3}{\milli\meter} \times \SI{25}{\micro\meter}$ Zr foil (\SI{99.2}{\percent} purity) and evaporated to dryness. Each drop contained \num{1.10(2)e15} Ho atoms. The sample was placed in a Ta tube (\SI{1.1}{\milli\meter} inner diameter, \SI{2}{\milli\meter} outer diameter). At vacuum conditions (about \SI{e-7}{\milli\bar}) in the ion-source region at temperatures of about \SI{560}{\degreeCelsius} the initial compound \ce{Ho(NO3)3*5H2O} is fully decomposed to \ce{Ho2O3} \cite{Balboul2000}. At about \SI{800}{\degreeCelsius} reduction of this oxide by the Zr foil sets in. Elemental Ho evaporates and diffuses to the ionization cavity where interaction with the laser radiation takes place \cite{Kieck2018}.

The ECHo $^{163}$Ho samples are in the same chemical form as the $^{165}$Ho AAS solution \cite{Dorrer2018}. In Section \ref{sec:conclusion} the isotope compositions of the initial samples are described. After dilution with MilliQ-water, the samples are handled the same way as described before. Therefore, all simulations and measurements are directly transferable to the case of $^{163}$Ho.

\subsection{Titanium-sapphire laser system}

For resonant two-step excitation, two lasers of the Mainz titanium-sapphire laser system were used in conventional Z-shaped resonator geometry \cite{Rothe2011}. The lasers operate in pulse mode at 10 kHz repetition rate and can be tuned from \SIrange{680}{960}{\nano\meter}, that almost completely cover the Ti:sapphire amplification range. Each of the lasers was pumped with \SIrange{15}{18}{\watt} of \SI{532}{\nano\meter} light from a commercial \SI{60}{\watt} frequency-doubled Nd:YAG laser. An average output power of \SI{3}{\watt} in the required range of fundamental wavelength was measured.

To overcome the low power of external single-pass second harmonic generation with values below \SI{300}{\milli\watt} for Ho \cite{Schneider2016}, the technique of intra-cavity frequency doubling \cite{Sonnenschein2015} was used. The efficient conversion leads to about \SI{500}{\milli\watt} output power of the second-step laser far beside the gain-maximum wavelength of the Ti:sapphire crystal. The second advantage of intra-cavity multi-pass second harmonic generation is a resulting high-quality beam profile without astigmatism, thus increasing the relative photon flux in the ionization volume.

The laser beams were focused and overlapped into the ionization cavity. By adjustment of the pump focus position within the Ti:sapphire crystal the 50 ns long output pulses were temporally synchronized and optimized for highest ion production rate, not for maximum laser power.

\subsection{Excitation scheme development}
\label{sec:scheme}

\begin{figure}[h]
	\includegraphics[width=0.48\textwidth]{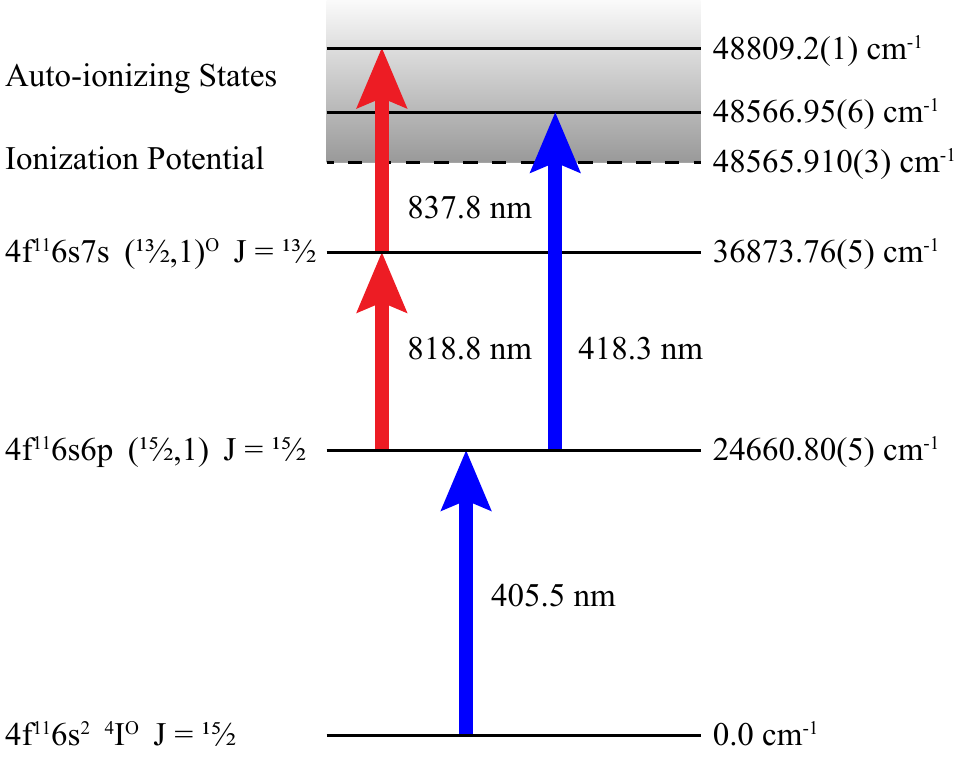}
	\caption{Comparison of the three-step and the newly developed two-step excitation scheme for Ho showing the corresponding atomic levels. The configurations and energies are taken from \cite{Kramida2018,Hostetter2015,Schneider2016} and this work.}
	\label{fig:ionization_scheme}
\end{figure}

In previous works on highly efficient laser ionization of Ho \cite{Gottwald2009,Liu2014,Schneider2016,Kieck2018}, a fully resonant three-step scheme was used, developed by Gottwald et al. and shown in Figure \ref{fig:ionization_scheme}. As it was pointed out by Schneider et al. \cite{Schneider2016} in detail, the third step does not exhibit saturation in the final excitation step even for the maximum accessible laser power. It reduces the excitation probability, but still leads to a quite high ionization efficiency of about \SI{40}{\percent} \cite{Liu2014,Schneider2016,Kieck2018}. Accordingly, the main goal for the ECHo project was the identification of an ionization scheme, for which full optical saturation and thus maximum possible ionization efficiency can be realized.

If $g_i$ is the degeneracy of the $i$th ionization scheme level, the partial level population of the final state $f$, as described in \cite{Letokhov1987}, is given by

\begin{equation}
	n_f = \frac{g_f}{\sum_{i}g_i}.
\end{equation}

Consequently, with fully saturated intermediate transitions and a sufficiently strong auto-ionizing transition, the ionization probability and efficiency for a two-step scheme is expected to be higher than for a three-step scheme.

\begin{figure}[h]
	\includegraphics[width=0.46\textwidth]{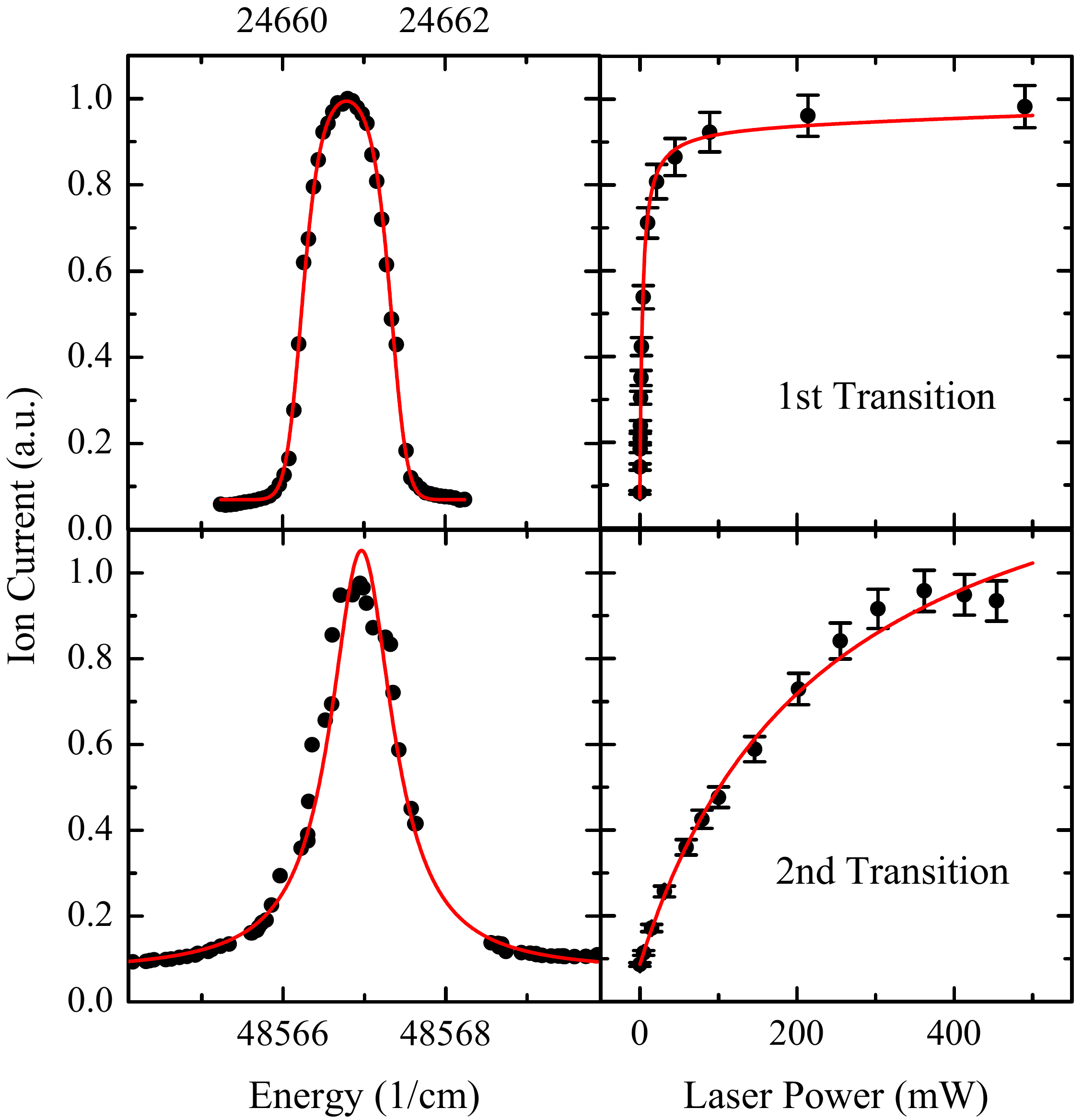}
	\caption{Spectral profile and saturation curve of the auto-ionizing transition. Results are discussed in the text.}
	\label{fig:2step}
\end{figure}

The enhanced output power in the second harmonic range and the higher spatial beam quality of the intra-cavity frequency-doubled titanium-sapphire laser system served to obtain full saturation in both transitions of a newly developed resonant two-step ionization scheme for Ho (Figure \ref{fig:ionization_scheme}). Both steps were characterized by studying the spectral profile and the laser-power dependence, both via recording the ion current. The measured data are shown in Figure \ref{fig:2step} and the fitting results are summarized in Table \ref{tabel:saturation}.

The first step was identical to that used as for the three-step scheme. The spectral profile of the first transition was measured at full laser power. The data was fitted with a power-broadened Gaussian function with the level of saturation $S$, the central wavenumber $\nu_c$, and the variance $\sigma^2$:

\begin{equation}
	I(\nu)=C_1\frac{1}{1+S^{-1}\exp(-\frac{(\nu -\nu_c)^2}{2\sigma^2})}+C_2.
\end{equation}

The second step was found by spectroscopic investigation of Ho and leads to a strong auto-ionizing state located just \SI[per-mode=reciprocal]{1}{\wavenumber} above the ionization potential (IP) of \SI[per-mode=reciprocal]{48565.910(3)}{\wavenumber} \cite{Hostetter2015}. The data are fitted with the theoretical model for the natural line shape of an auto-ionizing state from Fano \cite{Fano1961}:

\begin{equation}
	I(\epsilon)=\frac{(C+\epsilon)^2}{1+\epsilon^2}  , \ \ \epsilon(\nu)=\frac{\nu-\nu_c}{\Gamma/2}.
\end{equation}

\begin{table*}
	\centering
	\begin{tabular}{c S[table-format = 5.2(1),table-number-alignment = center] S[table-format = 5.2(1),table-number-alignment = center] S[table-format = 3.1(1),table-number-alignment = center] S[table-format = 1(1)e2,table-number-alignment = center]} 
		\toprule
		Transition & {$\nu_\textrm{lit}$} & {$\nu_\textrm{exp}$} & {$P_\textrm{sat}$} & {$C_\textrm{photo}$} \\ 
		& {(\si[per-mode=reciprocal]{\per\centi\meter})} & {(\si[per-mode=reciprocal]{\per\centi\meter})} & {(\si{\milli\watt})} & \\ 
		\midrule  
		1 & 24660.80(5) & 24660.8(4) & 3.3(3) & 7(2)e-5\\
		2 & {-} & 23906.16(3) & 240(80) & 0(3)e-4\\
		\bottomrule
	\end{tabular}
	\caption{Numerical results of the  line and saturation measurements of the two scheme transitions, shown in Figure \ref{fig:2step}. $P_\textrm{sat}$ is the saturation power of the transition and $C_\textrm{photo}$ the multiplier of the non-resonant photo-ionization.}
	\label{tabel:saturation}
\end{table*}

The spectral shape of the first transition is strongly power-broadened with $\textrm{FWHM}=\SI[per-mode=reciprocal]{1.11(5)}{\wavenumber}=\SI{33(2)}{\giga\hertz}$ width. This is much broader than the sum of about \SI{2}{\giga\hertz} Doppler broadening and \SI{6}{\giga\hertz} second-harmonic laser linewidth. The Fano profile of the second transition shows a width of $\textrm{FWHM}=\SI[per-mode=reciprocal]{1.12(3)}{\wavenumber}=\SI{34(1)}{\giga\hertz}$. 
 
The saturation is measured by gradually reducing the laser power with a variable attenuator. The data are shown in Figure \ref{fig:2step}. Both transitions were fitted to a laser-power dependent ion current model with constant background mainly caused by surface ionization (bg), a linear term for non-resonant photo-ionization (photo), and the saturation term for resonance ionization (sat):

\begin{equation}
	I(P) = C_\textrm{res} \frac{1}{1+P_\textrm{sat}/P} + C_\textrm{photo} P + C_\textrm{bg}.
\end{equation}

In Table \ref{tabel:saturation}, the negligible coefficient $C_\textrm{photo}$ confirms that the resonant photoionization is dominant over the non-resonant one. For the first transition, the power level $P_\textrm{sat}$ where half of the maximum ion current is reached, is about three times higher than in the measurements of Schneider et al. \cite{Schneider2016}. This can be caused by a broader spacial profile or rather a worse overlap of the atom cloud and the laser in the ionization region at the same laser power. Nevertheless, the first step is addressed with a laser power well above saturation. Unlike the three-step scheme, in the ionization steps of the new blue-blue scheme a saturation behavior is clearly observable. With the maximum accessible laser power of \SI{500}{\milli\watt} in the second step, about \SI{70}{\percent} of the ultimate ionization rate is reached. In addition, the saturated broad excitation profiles lead to the advantage, that both transitions tolerate a small drift in laser wavelength without measurable reduction in the efficiency.

\subsection{Ionization Efficiency}

\begin{figure*}[h]
	\centering
	\includegraphics[width=0.90\textwidth]{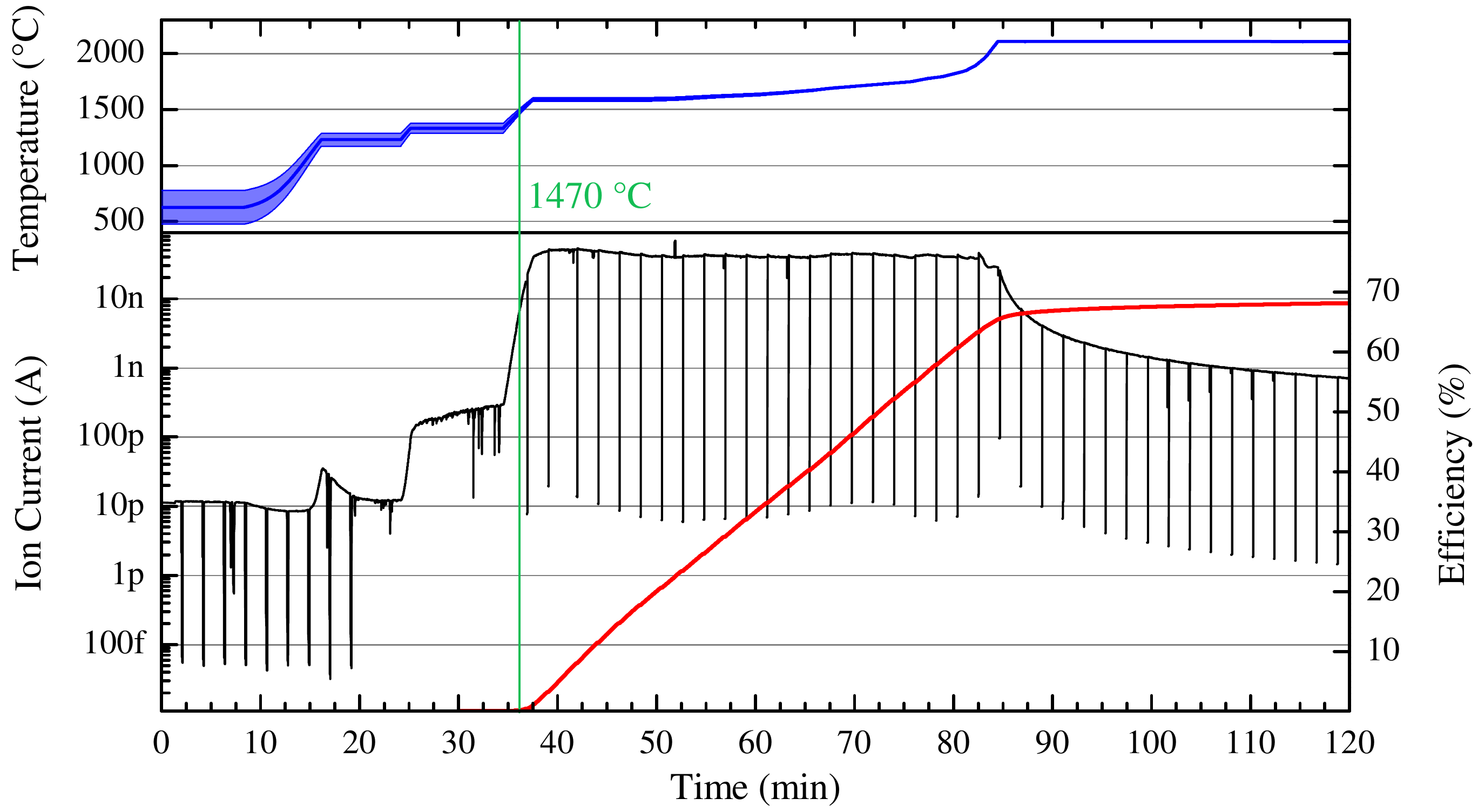}
	\caption{Time sequence of the fourth efficiency measurement from the current work, shown in Figure \ref{fig:efficiency}. Upper panel: sample temperature. Lower panel: ion current and cumulative efficiency. The vertical line indicates the time, where the melting point of Ho \cite{Spedding1960} was reached.}
	\label{fig:efficiency_example}
\end{figure*}

The characterization of the new two-step excitation scheme for Ho was done via a measurement of the achievable ionization efficiency. For this purpose a well-defined amount of Ho atoms (cf. Section \ref{sec:material}) is ionized and guided through the mass separator. The ion current is measured in a Faraday cup directly after the separator slit. With a ring electrode before the cup operated at \SI{-100}{\volt}, most of the secondary electrons from ion bombardment of the collecting electrode are repelled for a reliable ion current measurement. In Figure \ref{fig:efficiency_example} an example measurement is shown, where the respective temperature of the sample is indicated. The ionization cavity is gradually heated to constant operation condition with temperatures from \SIrange{1500}{2100}{\celsius}, measured between coldest and hottest region, respectively. The sample reservoir is then heated to a value of about \SI{1400}{\celsius}, where the ion current is high enough for laser and ion optics optimization. When the melting point of Ho of \SI{1470}{\celsius} \cite{Spedding1960} is exceeded, the ion current rapidly increases. For emulation the phase of detector implantation during the measurement, the ion current is stabilized between \SIlist{50;100}{\nano\ampere} by adaptively increasing the reservoir heating current and consequently controlling the sample temperature. After a period of \SI{45}{\minute}, over \SI{95}{\percent} of the final efficiency is reached and an ion current drop by more than one order of magnitude is observed. A further temperature increase was not carried out here, since it has no influence on the current decrease. With the sample slowly being fully exhausted, final increase of the last percentages of efficiency takes some hours.

\begin{figure}[h]
	\centering
	\includegraphics[width=0.46\textwidth]{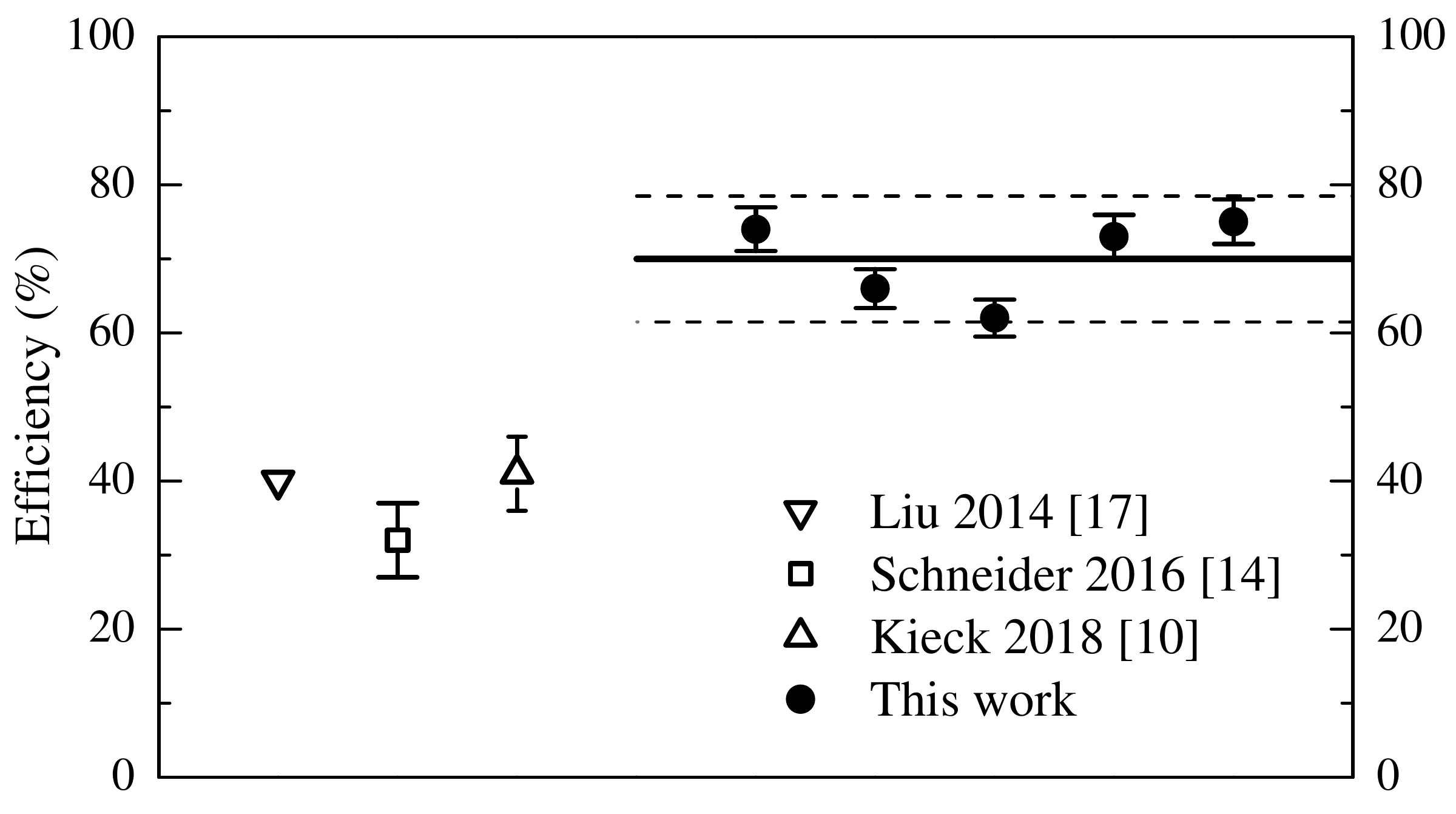}
	\caption{Efficiency measurements on Ho with two-step ionization scheme compared to previous measurements with three-step scheme. The mean value of the two-step scheme is \SI[parse-numbers=false]{69(5)\textsubscript{stat}(4)\textsubscript{sys}}{\percent}, as indicated by the horizontal lines. }
	\label{fig:efficiency}
\end{figure}

Before each efficiency measurement a blank Zr foil is measured in the same way to ensure that the ion source and the foil is not contaminated by stable Ho or any other interference on mass \SI{165}{\atomicmassunit}, stemming e.g. from memory effects of a previous run. Five efficiency measurements were performed and show results in very close agreement. The Figure \ref{fig:efficiency} compares the data with previous measurements on Ho performed with three-step resonance ionization \cite{Liu2014,Schneider2016,Kieck2018}, which resulted in significantly lower values. The error bars are determined by the inaccuracy of the sample preparation as described in Section \ref{sec:material}. The mean value of \SI{69}{\percent} is instrumentally weighted by the measurement errors: $w_i = 1 / \Delta x^2$. We define the statistical error of the efficiency measurement as the fluctuation between individual measurements that is well described by the mean absolute deviation $v$:

\begin{equation}
	v = \frac{\sum_{i} w_i | x_i - \bar{x} |}{\sum_{i} w_i}.
	\label{eq:deviation}
\end{equation}

Earlier measurements additionally indicated a systematic error occurring during the current recording of the \SI{30}{\kilo\electronvolt} ion beam in a Faraday cup because of secondary-electron and possible secondary-ion emission by sputtering. This leads to either an over- or underestimation of the ion current, respectively. We assess this systematic error with \SI{5}{\percent} of the measured efficiency. The new data confirm the expectations described in Section \ref{sec:scheme} about enhanced efficiency of a two-step ionization scheme over the previously employed three-step scheme. An increase of almost a factor of two results.

\subsection{Element selectivity}

The elemental selectivity is another advantage of resonant laser ionization compared to processes using high temperatures or electron/ion bombardment for ionization. During identification of suitable ionization schemes, the excitation steps must be monitored to ensure isobars of the isotope of interest are not ionized. If this condition is met, a contamination of the ion beam with other isotopes will mainly be caused by three effects:

\begin{itemize}
	\item Thermal ionization of isobaric atoms or molecules on the hot surfaces needed for the production of single atom vapor.
	\item Electron impact ionization of atoms or molecules induced by electron emission in high electric fields, e.g., in the extraction gap.
	\item Insufficient mass separation.
\end{itemize}

\begin{figure}[h]
	\centering
	\includegraphics[width=0.49\textwidth]{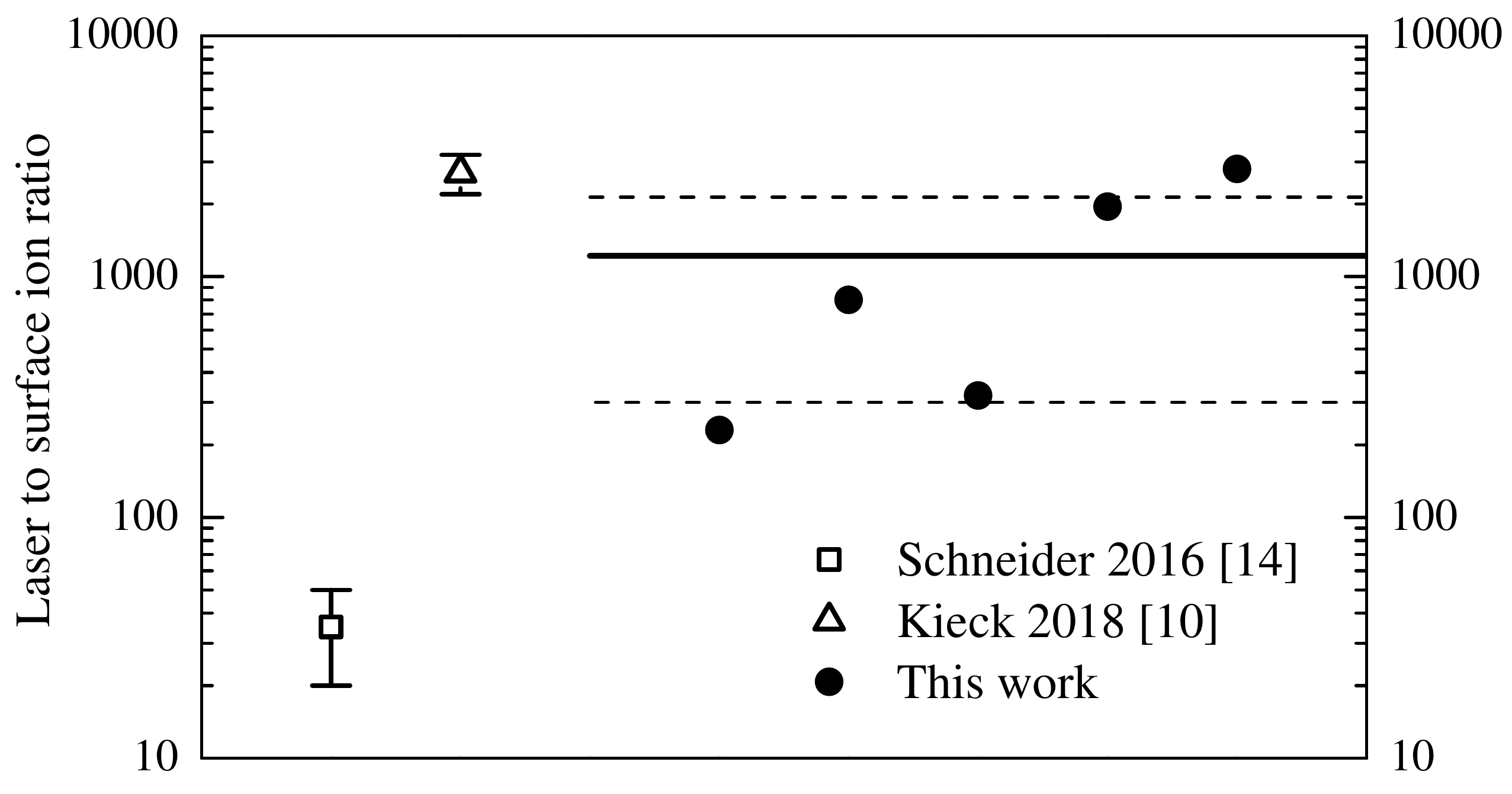}
	\caption{Laser to surface ion ratio measurements on Ho compared to previous measurements. The mean value is \num{1200(900)}, as indicated by the horizontal lines.}
	\label{fig:selectivity}
\end{figure}

While the mass separation of Ho isotopes is strongly influenced by the laser ionization as discussed in more detail in Section \ref{sec:separation}, a discrimination of the three interfering effects is neither necessary nor possible. Correspondingly, any interfering contamination of the ion beam will be termed background signal and is measured by repeatedly blocking the laser beams. This is responsible for the sharp drops in signal intensity every \SI{2}{\minute}, visible in Figure \ref{fig:efficiency_example}. The ratio of laser to background signal for the five efficiency measurements is calculated by dividing the accumulated ion charge by an integrated curve of the interpolated ion current extracted from these points. The results are given in Figure \ref{fig:selectivity} in comparison to previous measurements. The mean value of \num{1200} for signal to background is calculated using statistical weights $w_i = 1 / x_i$. It is about a factor of 2 lower than the value measured with the three-step scheme. With Equation \ref{eq:deviation} a high mean absolute deviation of \num{900} is calculated due to the substantial scatter of the data points. Additionally, it indicates a trend to higher values. As the data points in Figure \ref{fig:selectivity} are plotted in chronological order of their measurement, this could hint of the influence of the optimization procedure applied along the measurements. The average of the last two data points is \num{2400(400)}, which is identical within the experimental uncertainties with the three-step-scheme measurements taken in 2018.

\section{Ho isotope separation}
\label{sec:separation}

\begin{figure}[h]
	\centering
	\includegraphics[width=0.46\textwidth]{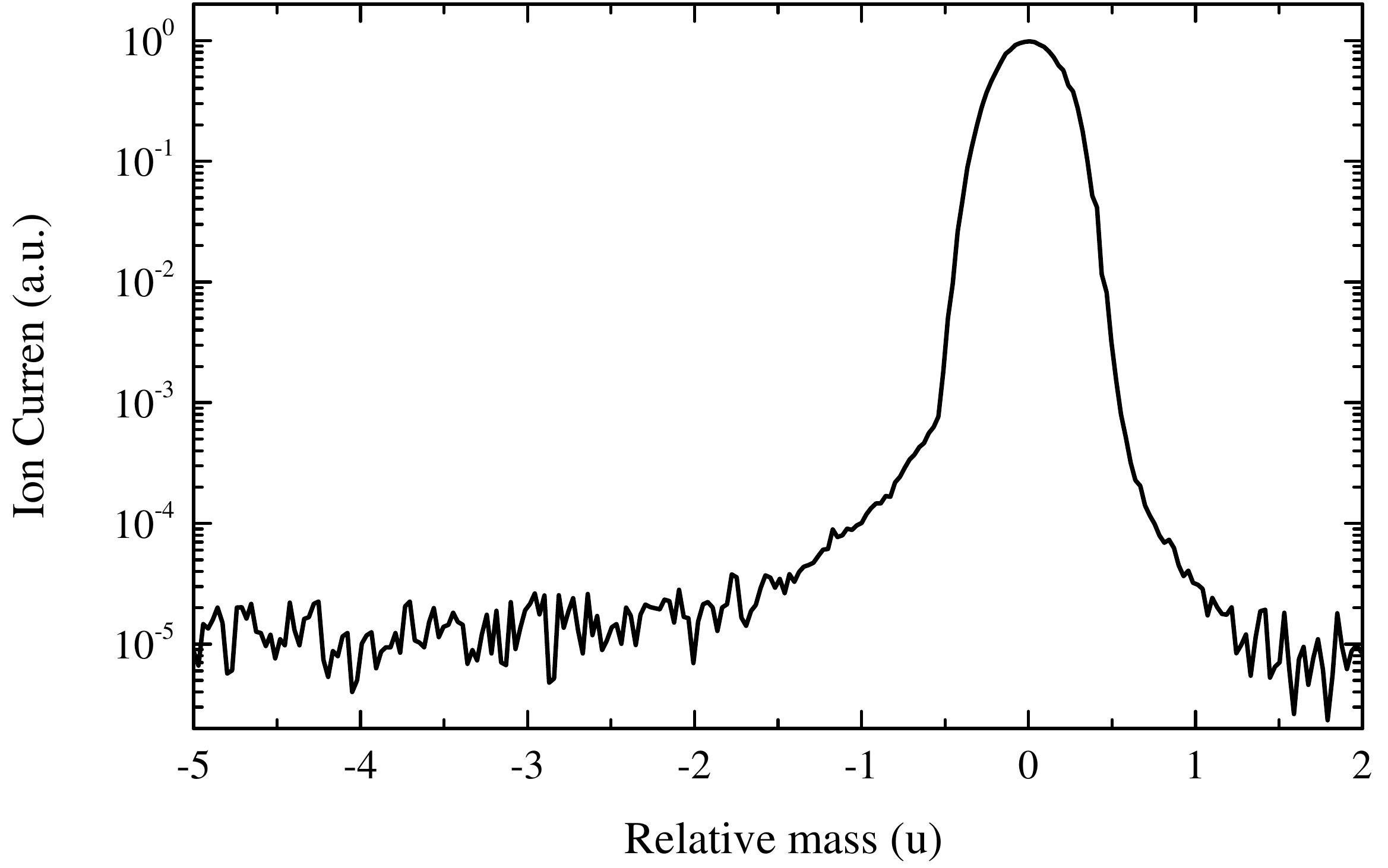}
	\caption{Mass scan performed with $^{165}$Ho in logarithmic scale. The separation factor of $^{166\textrm{m}}$Ho in the $^{163}$Ho implantation can be observed in the signal level at \SI{-3}{\atomicmassunit}.}
	\label{fig:masspeak}
\end{figure}

The elementally pure ion beam from the resonance ionization process is mass separated in a \SI{60}{\degree} double-focusing sector field magnet with a mass resolution of $m / \Delta m \geq 400$. An astigmatism of the ion beam caused by manufacturing tolerances of the magnet is adjusted by a quadrupole lens operated at \SI{20}{\volt} difference between horizontal and vertical electrodes. At the magnet focal plane, the desired mass is separated from its neighbors by a slit of about \SI{1}{\milli\meter} width, reducing the transmission by less than \SI{5}{\percent}. While the mass resolution primarily describes the central Gaussian part of the mass peak, the quantity is not suitable to calculate isotope separation factors with mass difference of \SI{3}{\atomicmassunit} or more. For recording an entire peak in the mass spectrum, the ion current of stable $^{165}$Ho is measured in a Faraday cup by an electrometer with high dynamic range, installed directly behind the separation slit and recorded while slowly scanning the magnetic field. A convolution of the mass peak and a top-hat function of the slit transmittance results (cf. Figure \ref{fig:masspeak}). This represents an integral of the ion current over the slit width in every data point. The tailing towards lower mass numbers is caused by resonant laser ionization of atoms present between ion source and extractor (cf. Figure \ref{fig:risiko}). This leads to a signal contribution at lower kinetic energy, i.e., stronger deflection in the magnet. 

Expectedly, $^{166\textrm{m}}$Ho is the dominant contamination in the ion beam before the separator magnet. The peak maximum at relative mass \SI{0}{\atomicmassunit} could be interpreted as its output of the ion source. Correspondingly \SI{3}{\atomicmassunit} lower would be the region to which the magnet must be tuned for implantation of $^{163}$Ho and the relative current level at this position would give the respective level of isotopic contamination. From \SI{-2}{\atomicmassunit} to still lower masses, the signal fluctuates at a level of about \SI{10}{\femto\ampere}. The mean value, in the range \SIrange{-2}{-5}{\atomicmassunit} is used for calculation of the suppression, which is defined as the separation factor. In this way RISIKO is shown to reduce the $^{166\textrm{m}}$Ho and other contaminants with mass \SI{166}{\atomicmassunit} and higher in the $^{163}$Ho implant to \num{\leq 1.4(6)e-5} of the amount before the magnet. The suppression of $^{165}$Ho, shown at \SI{-2}{\atomicmassunit}, i.e., in the lower-mass tailing of the peak is \num{1.0(5)e-4} of the initial amount. For suppression of lower masses on a higher mass position the separation is significantly better because of the absence of a high-mass tailing. The separation factor for a relative mass-difference of \SI[retain-explicit-plus]{+1}{\atomicmassunit} and \SI[retain-explicit-plus]{\geq +2}{\atomicmassunit} is \num{3(1)e-5} and \num{\leq 8(5)e-6}, respectively.

\section{Direct ion implantation}
\label{sec:implantation}
\subsection{Implantation setup}
\label{subsec:setup}
\begin{figure}[h]
	\centering
	\includegraphics[width=0.40\textwidth]{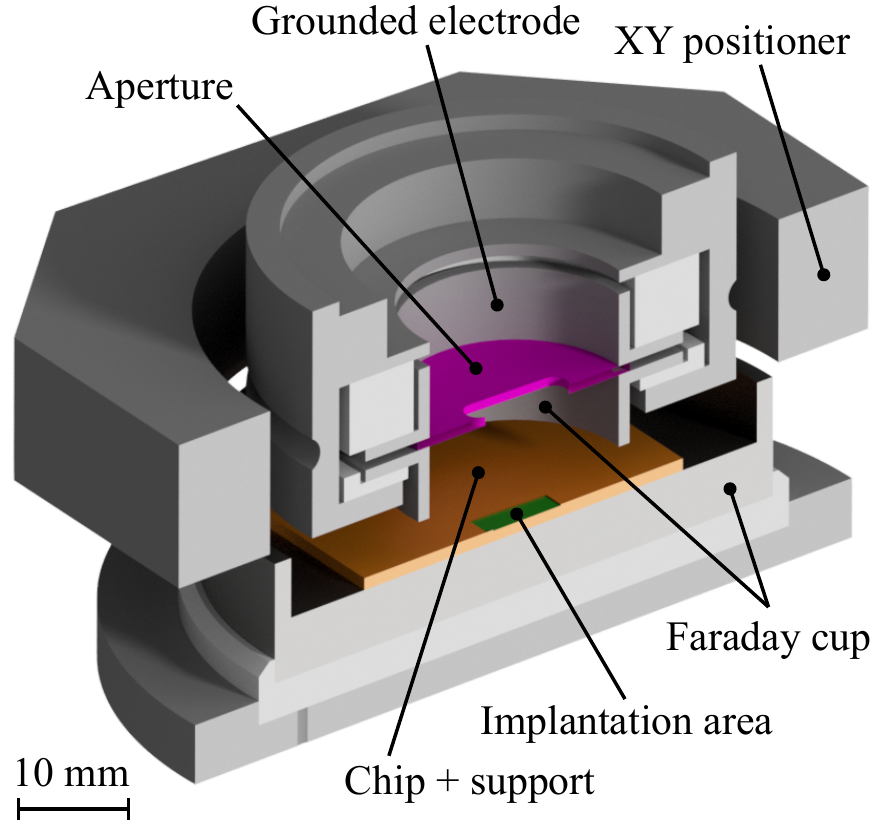}
	\caption{Setup for selective ion implantation into the calorimeter absorbers.}
	\label{fig:implantation_fixture}
\end{figure}

The ECHo \SI{10 x 5}{\milli\meter} detector chips is composed of an array of 64 metallic magnetic calorimeters (MMCs) with Au absorbers on top, \SI{180 x 180}{\micro\meter} each \cite{Gastaldo2017}. Prior to $^{163}$Ho ion implantation, the array must be prepared to protect the substrate and in particular the sensitive superconducting circuits from being destroyed by the impinging $^{163}$Ho ions and possible electric charging up. For that purpose they are covered with a few \si{\micro\meter} of photoresist. In the center of every absorber, an implantation area of \SI{160 x 160}{\micro\meter} is left uncovered. The top surface of the chip and chip support is then covered with a \SI{100}{\nano\meter} gold layer. For implantation, it is conductively fixed inside a Faraday cup to measure the ion current (cf. Figure \ref{fig:implantation_fixture}). To exclusively implant into the preselected area of absorbers on the chip, a \SI{8 x 3}{\milli\meter} aperture is placed in the flight path of the ions about \SI{9}{\milli\meter} in front of the implantation spot, as shown in Figure \ref{fig:implantation_fixture}. The rear surface of this aperture is isolated from the grounded outer parts and kept in electric contact to the Faraday cup interior in a way ensuring that most secondary electrons and ions cannot escape and falsify the ion current measurement.

The entire aperture stack is mounted in a rotatable XY positioner with micrometer adjustment. For accurate positioning of the MMC arrangement and aperture, a \SI{1}{\milli\watt} laser beam is introduced in perpendicular geometry onto the implantation area. Subsequently, the aperture mount is installed and positioned precisely relative to the laser beam. This procedure allows a position accuracy in the order of \SI{200}{\micro\meter}.

\subsection{Geometric implantation efficiency}
Behind the beam waist, which must be set into the separation slit of the magnet (cf. Figure \ref{fig:risiko}), the ion beam is diverging. In the implantation region, the beam diameter would thus be about \SI{5}{\milli\meter} (FWHM). In this way only \SI{9}{\percent} of the $^{163}$Ho ions would implant into the absorber array, while the majority would be incorporated into the surrounding gold layer and the aperture. To enhance the low implantation efficiency, a dedicated Einzel lens together with an electrostatic XY deflector was installed between separation slit and implantation region. Operated at values around \SI{14}{\kilo\volt}, this arrangement leads to a secondary focal plane at the implantation position on the chip. In addition, the properly focused ion beam is frequently scanned across the surface area of the detector chip to cover all 64 absorbers.

\begin{figure*}[h]
	\centering
	\includegraphics[width=1\textwidth]{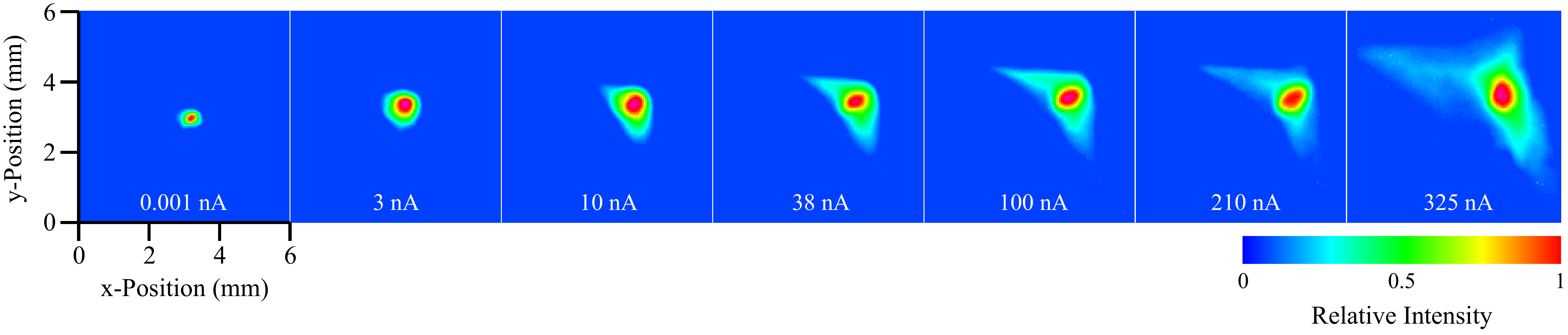}
	\caption{MCP images of the focused ion beam at different ion currents with the same position scaling. The color profile is normalized to the maximum intensity within each of the seven shown profiles. Color version online.}
	\label{fig:beam_characteristics}
\end{figure*}

To characterize the performance of this arrangement and a possible negative influence by space charge effects, the focusing capability of the ion beam was measured by imaging the beam as function of the ion current from \SIrange{0.001}{325}{\nano\ampere}. The ion beam density profile was analyzed using a chevron stack of two micro channel plates with an adjacent phosphorous screen (MCP). Images were recorded by a digital camera with an intensity resolution of 10 bit. The spacial resolution of the whole configuration is \SI{50 x 50}{\micro\meter}. In Figure \ref{fig:beam_characteristics} the intensity profiles are shown as a function of the ion current. The maximum current is limited by the MCP to values significantly lower than those that can be transmitted by the RISIKO mass separator (\SI{>1}{\micro\ampere}).

With increasing ion current, the two-dimensional intensity distribution starts to deviate from a circular profile as low-intensity wings start to appear. These regions do not contribute substantially to the total transmitted ion current. However, the horizontal wing extends towards lower x-positions, where ions of lower magnetic rigidity, hence lower mass, are present. This would be most critical in case of $^{166\textrm{m}}$Ho ions, where this low energy wing could extend into the beam area of $^{163}$Ho and be co-implanted. In the sample, the amount of $^{166\textrm{m}}$Ho is about four orders of magnitude lower than the one of $^{163}$Ho \cite{Dorrer2018}. Hence, the ion beam current of $^{166\textrm{m}}$Ho will be well below \SI{1}{\nano\ampere}, in which case no appreciable wing extending into the $^{163}$Ho region is expected. The vertical extend of the ion beam profile is also reproducible and could be caused by small manufacturing defects of the magnet pole pieces. It does not affect the beam purity but lowers the implantation efficiency at high ion currents. The beam shape and dimensions remain almost constant within the range of \SIrange{3}{210}{\nano\ampere}, assuring identical characteristics for ion implantation. In this range, the implantation efficiency is \SI{20(2)}{\percent}. At \SI{325}{\nano\ampere}, the dimension increases, as the system limitation is reached.

\section{Conclusion}
\label{sec:conclusion}
The RISIKO mass separator at the Johannes Gutenberg University (JGU) Mainz has been upgraded and optimized for ion implantation of $^{163}$Ho into the detector array of the ECHo experiment for neutrino mass determination. Special emphasis was placed on achieving the highest efficiency and beam purity. The enhanced output power of the intra-cavity doubled Ti:sapphire laser allowed obtaining high saturation in both transitions of the newly-developed two-step resonance ionization scheme. A high efficiency of \SI[parse-numbers=false]{69(5)\textsubscript{stat}(4)\textsubscript{sys}}{\percent} has been reproduced in a series of five measurements using samples with \num{1.10(2)e15} atoms of stable Ho. To our knowledge this is presently the highest overall efficiency obtained with a laser ion sources, which is in the same range as data for Pd measured at JGU and ORNL \cite{Kron2016}. The ion source can be accurately controlled to deliver a stable output current in the order of \SI{100}{\nano\ampere}, by which the sample is depleted within a few hours.

The combination of resonance ionization and high-transmission magnetic sector field mass separation will increase the purity of the implanted $^{163}$Ho source to at least \SI{99.9992(4)}{\percent}. Using starting materials with a $^{166\textrm{m}}$Ho:$^{163}$Ho ratio of \num{2.4(3)e-5} and \num{3.2(6)e-4} \cite{Dorrer2018}, this will lead to a $^{166\textrm{m}}$Ho contamination in the MMCs below \num{3(1)e-10} and \num{4(2)e-9}, respectively. In Table \ref{tabel:composition} the abundance of all relevant isotopes are calculated for every separation step.

\begin{table*}[h]
	\caption{\label{tabel:composition}Calculated composition change of the two analyzed ECHo $^{163}$Ho samples through resonance ionization (RI) and mass separation (MS). The initial composition is taken from \cite{Dorrer2018}.}
	\centering
	\begin{tabular}{c @{\hskip 3\tabcolsep} r @{\hskip 3\tabcolsep} S[table-format = 1.1(1)e+2,table-number-alignment = center] @{\hskip 3\tabcolsep} S[table-format = 1.1(1)e+2,table-alignment = center] @{\hskip 3\tabcolsep} S[table-format = 3.1(1)e+2,table-alignment = center] @{\hskip 3\tabcolsep} l @{\hskip 0.4\tabcolsep} l  @{\hskip 0.4\tabcolsep} l}
		\toprule
		Sample from & Nuclide & {Initial} & {Ionized} 		& {Implanted}			&\multicolumn{3}{@{}l@{\hskip0.4\tabcolsep}}{Method}\\ 
		\midrule 
		\multirow{8}{*}{\SI{6.7}{\milli\gram} Er target} & $^{160}$Dy & 4(1)e-5 & 2(1)e-8 & {$\leq$}2(1)e-13 & RI&+&MS \\  
		& $^{161}$Dy 	& 3(1)e-4		& 1.3(3)e-7 		& {$\leq$}1.2(8)e-12 	& RI&+&MS \\
		& $^{162}$Dy 	& 7(1)e-4		& 3(1)e-7	 		& 1.0(4)e-11 			& RI&+&MS \\
		& $^{163}$Dy 	& 7(1)e-4		& 3(1)e-7	 		& 4(1)e-7				& RI&& \\
		& $^{164}$Dy 	& 6(1)e-4		& 3(1)e-7	 		& 4(2)e-11				& RI&+&MS \\
		& $^{163}$Ho 	& 0.887(6)		& 0.889(6)			& 0.9999978(8)			& & & \\
		& $^{165}$Ho	& 0.111(6)		& 0.111(6)			& {$\leq$}2(1)e-6		& & &MS \\
		& $^{166\textrm{m}}$Ho & 2.1(2)e-5& 2.1(2)e-5		& {$\leq$}3(1)e-10		& & &MS \\
		\midrule
		\multirow{14}{*}{\SI{30}{\milli\gram} Er target}& $^{160}$Dy & 2(1)e-5 & 2(1)e-8 & {$\leq$}2(1)e-13 & RI&+&MS \\  
		& $^{161}$Dy 	& 2(1)e-4		& 8(2)e-8	 		& {$\leq$}1.0(7)e-12 	& RI&+&MS \\
		& $^{162}$Dy 	& 3(1)e-4		& 2(1)e-7	 		& 7(3)e-12 				& RI&+&MS \\
		& $^{163}$Dy 	& 6(1)e-4		& 3(1)e-7	 		& 4(1)e-7				& RI && \\
		& $^{164}$Dy 	& 3(1)e-3		& 1.4(3)e-6 		& 2(1)e-10				& RI&+&MS \\
		& $^{163}$Ho 	& 0.59(5)		& 0.66(5)			& 0.999992(4)			& & & \\
		& $^{165}$Ho	& 0.31(4)		& 0.34(5)			& {$\leq$}7(4)e-6		& & &MS \\
		& $^{166\textrm{m}}$Ho& 1.9(2)e-4& 2.1(3)e-4		& {$\leq$}4(2)e-9		& & &MS \\
		& $^{162}$Er	& 1.7(4)e-2		& 8(2)e-6	 		& 4(2)e-10				& RI&+&MS \\  
		& $^{164}$Er 	& 4(1)e-3		& 2(1)e-6	 		& 3(2)e-10		 		& RI&+&MS \\
		& $^{166}$Er 	& 3(1)e-2		& 1.4(4)e-5 		& {$\leq$}3(2)e-10		& RI&+&MS \\
		& $^{167}$Er 	& 2.1(4)e-3		& 1.0(3)e-7 		& {$\leq$}2(1)e-11		& RI&+&MS \\
		& $^{168}$Er 	& 4(1)e-2		& 2(1)e-5	 		& {$\leq$}4(2)e-10		& RI&+&MS \\
		& $^{170}$Er 	& 7(2)e-3		& 3(1)e-6			& {$\leq$}7(4)e-11		& RI&+&MS \\
		\bottomrule
	\end{tabular}
\end{table*}

With the newly developed ECHo implantation setup, an ion beam up to \SI{210}{\nano\ampere} can be handled and accurately implanted. A geometric efficiency of \SI{20(2)}{\percent} has been extracted for the typical arrangement of the ECHo detector array. With the current sample amount of \num{1.2(2)e18} atoms of $^{163}$Ho after the chemical separation \cite{Dorrer2018}, this would deliver \num{1.7(4)e17} atoms in the detectors by applying ionization, separation and geometric implantation efficiency, which corresponds to an activity of about \SI{800(200)}{\kilo\becquerel}. This is in the range requested for the presently running project phase of ECHo 100k \cite{Gastaldo2017}. With ion currents of \SI{30}{\nano\ampere}, the implantation of an ECHo chip with \SI{10}{\becquerel} in each of the 64 absorbers takes about \SI{1}{\hour}.

The abundance of the interfering $^{166\textrm{m}}$Ho in relation to $^{163}$Ho in the implant meets the requirements of a value below \num{5e-9} \cite{Gastaldo2017}. This, however, should be verified in independent studies using Accelerator Mass Spectrometry (AMS) or comparable techniques capable of determining extreme isotope ratios.

\section{Acknowledgments}
This work was performed under the framework of the DFG Research Unit FOR 2202 (ECHo) with financial support under the contracts DU 1334/1-1 and DU 1334/1-2.

\FloatBarrier
\section*{References}
\bibliography{mybibfile}

\end{document}